\documentclass[a4paper,11pt,twoside]{article}
\pdfoutput=1
\usepackage{amsmath,amssymb}
\usepackage{paralist}
\usepackage{slashed}
\usepackage{graphicx}
\usepackage{subfigure}
\usepackage{color}
\usepackage{fullpage}
\usepackage{multirow}

\usepackage[]{hyperref}

\title{
\vspace{-3cm}
\begin{flushright}
\small{CERN-PH-TH/2012-107}
\end{flushright}
\vspace{3cm}
\bf \huge
Implications of a Light Higgs\\ in Composite Models
\vspace{.2cm}}
\date{}
\author{
{\large  Michele Redi$^{a,b}$\footnote{michele.redi@cern.ch},~
Andrea Tesi$^{c}$\footnote{andrea.tesi@sns.it}}\\
[10mm]
\normalsize\itshape
$^a$ CERN, Theory Division, CH-1211, Geneva 23, Switzerland\\
\normalsize\itshape $^b$ INFN, Sezione di Firenze, Via G. Sansone, 1; I-50019 Sesto Fiorentino, Italy
\\
\normalsize\itshape
$^c$ Scuola Normale Superiore, Piazza dei Cavalieri, 7; I-56126 Pisa, Italy}

\begin{document}
\maketitle \thispagestyle{empty}
\begin{abstract}
\noindent
We study the Higgs mass in composite Higgs models with partial compositeness,
extending the results of Ref. \cite{4dcomposite} to different representations of the composite sector for $SO(5)/SO(4)$
and to the coset $SO(6)/SO(5)$. For a given tuning we find in general a strong correlation between the mass of the top partners and the Higgs mass,
akin to the one in supersymmetry. If the theory is natural a Higgs mass of 125 GeV typically requires fermionic partners below TeV which
might be within the reach of the present run of LHC. A discussion of CP  properties of both cosets is also presented.
\end{abstract}

\newpage
\setcounter{page}{1}
\section{Introduction}
\label{sec:1}

We are likely at the dawn of the discovery of a light Higgs boson \cite{atlas,cms}.
A pressing question is whether the Higgs is Standard Model-like, or if there is new dynamics that stabilizes the electro-weak scale.

Two scenarios are at present the most compelling solutions of the hierarchy problem: supersymmetry and strong dynamics where the Higgs
is a bound state of new strong interactions \cite{georgi,minimalcomposite}. In the first at least new colored scalars,
the stops, are expected below the TeV scale if the theory shall remain natural, while other supersymmetric partners
could be heavier. If strong dynamics in the form of a Composite Higgs
(CH) boson  is responsible for electro-weak symmetry breaking, we also expect new resonances around the TeV scale,
with the same statistics as the SM fields in this case. But how heavy? This quantitative question is obviously
crucial for the prospect of detecting the new states at the LHC. As we will see in a large class of models
a conclusion similar to SUSY applies even quantitatively.
Namely,  if a light Higgs will be discovered, for a similar tuning as in SUSY new fermionic resonances must be present below
the TeV scale to cut-off the top loop quadratic divergence. Vector resonances could instead be in the multi-TeV range,
as hinted by precision tests, $S$ parameter in particular. In other words a hierarchical spectrum is predicted.

In this note we will focus on models where the Higgs is a pseudo-Goldstone boson as this is the only clear logic,
beside supersymmetry, that allows one to obtain a naturally light scalar, see \cite{continoreview} for a review.
Two patterns stand out phenomenologically, the minimal composite $SO(5)/SO(4)$ \cite{minimalcomposite}
delivering a single Higgs boson and $SO(6)/SO(5)$ that in addition produces a CP odd singlet \cite{singlet}. Extended Higgs sectors
are also possible \cite{cthdm} but we will not consider them here. Within this class we will consider scenarios with partial compositeness where
the Higgs potential is generated by the couplings to the SM fields that explicitly break the global symmetries of the strong sector.

To address quantitatively the question of the mass  we will use the simplified description of CH models with partial compositeness
developed in \cite{4dcomposite} (see \cite{discretecomposite} for an alternative construction and \cite{contino-sundrum,SILH,wacker,thaler} for related work).
In this framework only the lightest resonances of the strong sector are introduced allowing a model independent analysis.
This turns out to be sufficient to render the potential calculable. The results are comparable to the ones of 5D models for similar choices of
parameters but other regions  of parameter space are also explored. While not true in general, in all the models considered the top Yukawa
controls the Higgs mass so that a relation between the fermionic partners and the tuning is obtained analogous to the one
in supersymmetry for the stops.

The paper is organized as follows. In section \ref{sec:2} after  briefly reviewing the simplified approach of Ref. \cite{4dcomposite},
we study the Higgs and fermionic partners mass in the minimal coset  $SO(5)/SO(4)$ with fermions in the \textbf{5} and in the \textbf{10} representations.
We also consider the possibility, motivated by flavor, that right-handed light quarks are strongly composite which generates new contributions
to the potential but with similar results for the spectrum. Section \ref{sec:3} is devoted to the study of the coset space $SO(6)/SO(5)$.
The mass of the Higgs is similar to the minimal CH while the singlet is typically heavier.
We investigate in detail the CP properties of this model showing that the CP symmetry is not broken spontaneously nor explicitly
by the coupling to the SM fermions. Conclusions and a discussion of the analogies with supersymmetry is in section \ref{sec:4}.
The relevant formulas used in the paper are collected in appendix \ref{appA}. Some  analytic formulas and intuition for the potential are
in appendix \ref{sec:B}.

\section{Coset Space $SO(5)/SO(4)$}
\label{sec:2}

In Ref. \cite{4dcomposite} we presented an effective lagrangian that encodes all the relevant features
of CHM with partial compositeness and contains only the relevant degrees of freedom
possibly accessible at the LHC (see also Ref. \cite{discretecomposite}).
Our simplified model contains for each SM fermion a single Dirac fermion in a representation of the global symmetry
of the composite sector. As in the 5D models SM chiral fermions are associated to independent multiplets.
Within this framework many observables of interest are calculable.
In particular the Higgs potential, UV divergent in the low energy effective theory, is finite due to the presence of the resonances
and can be used to estimate the Higgs mass up to model dependent effects.
This reproduces results similar to 5D models for analogous choices of parameters.

The construction can be applied to any CHM. We start considering the simplest coset $SO(5)/SO(4)$
with the most popular choices of the embedding of fermions, the $\bf{5}$ and the $\bf{10}$. This pattern of symmetry
breaking was studied in \cite{4dcomposite} and we refer to that paper for the technical details.
We collect some of the relevant formulas in appendix \ref{appA}. Approximate analytic formulas
for the potential are found in appendix \ref{sec:B}.

\subsection{Gauge Lagrangian}

In the gauge sector the lagrangian reads \cite{4dcomposite},
\begin{equation}
\begin{split}
{\cal L}_{gauge}&=-\frac 1 {4 g^2_{0}} F^a_{\mu\nu} F^a_{\mu\nu}-\frac 1 {4 g^2_{0Y}} Y_{\mu\nu} Y^{\mu\nu} \\
&+\frac{f_1^2}{4}{\rm Tr}\left|D_{\mu}\Omega \right|^2 +  \frac {f_2^2} 2 \left(D_{\mu}\Phi\right)^T\left(D^{\mu}\Phi\right) -\frac 1 {4 g_\rho^2}
\rho_{\mu\nu}^A\rho^{A\mu\nu}
\end{split}
\label{gaugeminimal}
\end{equation}
where $\Omega$ is an $SO(5)$ matrix parametrizing the coset $SO(5)_L\times SO(5)_R/SO(5)_{L+R}$, and $\Phi$ is a five dimensional unit vector
whose VEV breaks spontaneously $SO(5)$ to $SO(4)$. The composite spin-1 resonances are introduced as gauge fields of the diagonal subgroup of $G_R+G$
and the SM fields gauge the electro-weak subgroup of the $SO(5)_L$ global symmetry of the composite sector,
\begin{equation}
D_{\mu}\Omega=\partial_{\mu}\Omega-i A_{\mu} \Omega+i \Omega \rho_{\mu},~~~~~~~D_{\mu}\Phi=\partial_{\mu}\Phi-i \rho_{\mu} \Phi
\end{equation}
where $A_\mu$ are the elementary gauge fields. The physical decay constant of the 4 GBs (the Higgs) is,
\begin{equation}
f^2=\frac {f_1^2 f_2^2}{f_1^2+f_2^2}
\end{equation}
To reproduce the hyper-charge assignments of fermions a $U(1)_X$ symmetry should also be included so that $Y=T^3_R+X$.

\subsection{CHM$_5$}
\label{secCHM5}

Each SM quark is coupled to a distinct Dirac fermion in an $SO(5)$ representation. The spontaneous breaking $SO(5)/SO(4)$
allows couplings between fermions associated the left and right chiralities of SM fields that will eventually generate SM Yukawas.
For the model where the fermions are in the ${\bf 5}$ of $SO(5)$ (CHM$_5$ \cite{custodian}) the lagrangian of the third generation reads,
\begin{equation}\label{CHM5}
\begin{split}
{\cal L}^{\rm CHM_5} &=  \bar{q}^{el}_L i \slashed{D}^{el} q^{el}_L +  \bar{t}^{el}_R i \slashed{D}^{el} t^{el}_R \\
&+ \Delta_{t_L}\,  \bar{q}^{el}_L \Omega \Psi_{T}  +\Delta_{t_R}\, \bar{t}^{el}_R \Omega \Psi_{\widetilde{T}} +  h.c. \\
 & + \bar{\Psi}_T (i \slashed{D}^{\rho} -m_{T}) \Psi_T +  \bar{\Psi}_{\widetilde{T}} (i \slashed{D}^{\rho} -m_{\widetilde{T}}) \Psi_{\widetilde{T}}   \\
 & -  Y_T \bar{\Psi}_{T,L} \Phi \Phi^T \Psi_{{\widetilde{T}},R}- m_{Y_T}\bar{\Psi}_{T,L} \Psi_{{\widetilde{T}},R}+h.c.\,  \\
 & + (T\rightarrow B).
\end{split}
\end{equation}
The elementary quarks $q_L^{el}$ and $t_R^{el}$ couple to Dirac fermions in the $\mathbf{5}$ of $SO(5)$ ($\Psi_{T(\widetilde{T})}$) via mass mixing $\Delta_{t_L}$ and $\Delta_{t_R}$ that respect the SM gauge symmetry. The terms in the fourth line break spontaneously $SO(5)/SO(4)$ and contain interactions with the GBs. We retain the only terms with a certain chirality as necessary to generate the SM Yukawas (see \cite{4dcomposite} for more details). We recall that in CHM$_5$ the SM quark doublet must couple to two composite fermions with different charge under $U(1)_X$
to generate Yukawa of the top and bottom quark. The couplings of the down sector (to which we refer in \eqref{CHM5} in the last line) however will not be important here. Within the anarchic scenarios \cite{minimalcomposite} the potential is dominated by the third generation because
the mixings of the light generations are small however this might not be true in general and we provide an example below.

In general $m_{Y_T}$ and $Y_T$ are complex parameters. One phase can be reabsorbed with a redefinition of the composite fields
while the relative phase remains as a physical CP violating phase. This describes a strong sector that breaks CP.
The same holds when the coupling to the elementary fields is included so that the action violates in general CP
even with a single generation. This phase will not appear in the Higgs couplings (see end of section \ref{sec:3})
but in more subtle observables such as correlation of masses. Following the literature we will take these IR parameters
to be real in what follows, i.e.  we consider a CP invariant composite sector.

\begin{figure}[t!]
\begin{center}
\subfigure
{\includegraphics[width=0.65\textwidth]{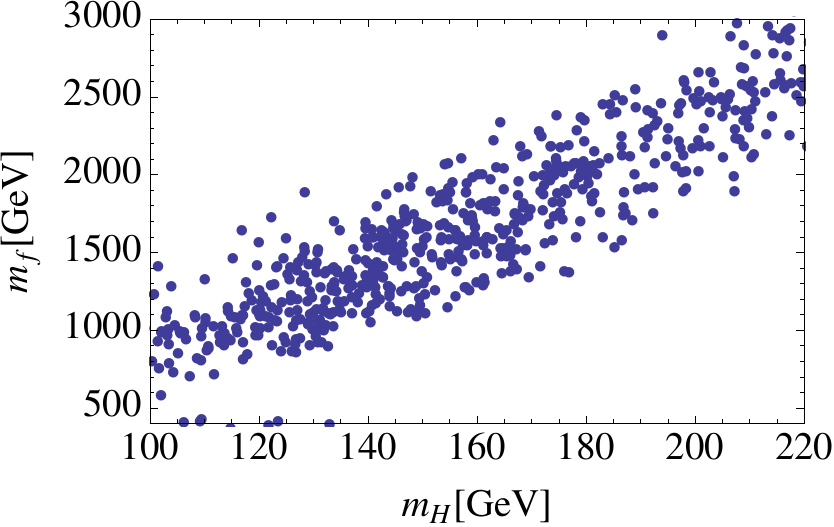}}\quad
\subfigure
{\includegraphics[width=0.65\textwidth]{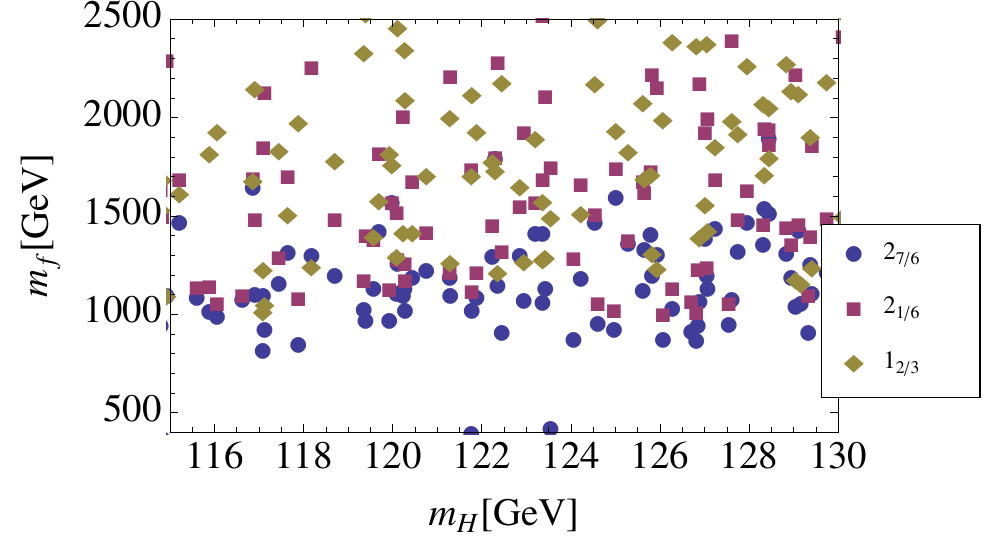}}
\caption{\label{fig:SO5Higgs-all}\small Masses of fermionic partners as a function of the Higgs mass for $f=800$ GeV in CHM$_5$.
The six fermionic parameters are varied between $0.3$ and $4$ TeV and we require mixing elementary composite $\Delta_{t_L,t_R}/m_{T,\tilde{T}}<3$.
The gauge contribution corresponds to $f_1=f_2=\sqrt{2}f$  and $g_\rho=3$. In the first plot mass of the lightest fermionic
partner as a function of the Higgs mass. In the second plot mass of the fermionic excitations in the low mass region.}
\end{center}
\end{figure}

In \cite{4dcomposite} we presented various scans for CHM$_5$. There we used as a benchmark $f=500$ GeV also in order to
compare with results obtained in 5D models. This value of $f$ leads to a certain
tension with electro-weak precision tests that typically requires new contributions (for example to the $T-$parameter) to agree
with the data. Moreover for a light Higgs, as hinted by recent LHC data, the masses of the fermionic resonances are often lighter than $500$ GeV
that is on the verge of being excluded by direct searches \cite{heavy-searches}. In this paper we present our results for the more conservative and perhaps more realistic choice $f=800$ GeV corresponding to a tuning parameter $\xi=v^2/f^2\simeq 0.1$, where precision tests are more easily satisfied\footnote{Our results can be easily extrapolated to different values of $f$. Neglecting higher order terms in $v^2/f^2$ and effects associated
to the running of couplings (corresponding to higher loop corrections) we can rescale $f$ and all other dimensionfull parameters of a given point and obtain in the vicinity of
this point the electro-weak VEV, approximately equal Higgs mass and fermion masses rescaled as $f$. This is because the Higgs mass is controlled by the dimensionless quartic coupling that is not rescaled. On the other hand the amount of cancellation of the quartic terms  will grow proportionally to $f^2$.}. For this value the spin-1 resonances can be easily above 3 TeV so that contributions to the $S-$parameter are sufficiently suppressed and model dependent contributions to $T$ are typically within experimental bounds. In this case the resonances are heavier but could still be within the reach of the early LHC. The Higgs couplings are within 10\% as in the SM so clearly our choice should be reconsidered if large deviations will be measured.  Indeed this possibility is not yet excluded by present results \cite{lightHiggs}.

As in our previous study we have performed a scan over the 6 fermionic parameters of the model,
requiring that the correct electro-weak VEV and top mass are generated.\footnote{We demand $m_t\in[145,155]$ to roughly account for the running of the top mass which is generated at the scale of the heavy fermions $\sim $ TeV. A detailed analysis of the RGE is beyond the scope of our work.}
We have improved our numerical procedure in several ways compared to \cite{4dcomposite}, including in particular all order corrections in $v/f$.
Our plot are obtained using the standard approximation for the potential,
\begin{equation}
\label{potMFV}
V(h)\approx  \alpha s_h^2 - \beta s_h^2 c_h^2\,,
\end{equation}
where $s_h=\sin (h/f)$ and $h$ is the physical Higgs scalar. We have also performed the computations with the exact 1-loop Coleman-Weinberg effective potential
finding that the corrections are negligible for $f=800$ GeV while they might be  important for lower values of $f$.
Requiring that the electro-weak VEV is reproduced one finds\footnote{The electro-weak VEV is defined as $v=f \sin (\langle h\rangle/f)$, where $v=246$ GeV.},
\begin{equation}
m_h^2\approx 8  \frac  {\beta}{f^4} v^2
\label{mhbeta}
\end{equation}
which depends only on $\beta$.  The coefficient $\alpha$ is proportional to the left or right mixings squared while $\beta$ is proportional to
the top Yukawa squared. From the low energy point of view these contributions can be understood as generated from the loops
of the Higgs dependent kinetic terms and of the top Yukawa respectively \cite{4dcomposite}.
The natural size of $\beta$ is,
\begin{equation}
\beta\approx \frac {N_c\, y_t^2}{16\pi^2} f^2 \Lambda^2
\label{betanatural}
\end{equation}
where $\Lambda$ is the cut-off entering the top loops physically represented by the fermionic resonances.
This is exactly reproduced in our model, see appendix \ref{sec:B}.
From this the degree of tuning $m_h^2/\delta m_h^2$ scales as $v^2/f^2$.  This can be considered as a lower bound on the tuning
that is often larger because the typical size of $\alpha$ is larger than $\beta$.

The result of the scan is reported in Figs. \ref{fig:SO5MFVHiggs-all}. In the first figure we show the correlation of the Higgs and the lightest
fermionic resonances which in a large fraction of points is the doublet of hypercharge 7/6 (the ``custodian''), even though
regions of parameters where the singlet is the lightest state can be found.
Splittings generated by electro-weak symmetry breaking are neglected throughout.
In the lower figure we zoom on the low Higgs mass region allowed by the LHC and show the mass of the singlet,
doublet and custodian fermions.

A comment is in order. The natural size $\alpha$ is larger than $\beta$ so tuning the electro-weak VEV requires
$\alpha\ll \alpha_{\rm{naive}}$. This is often ascribed to a cancellation between the contributions associated to the top left and top right.
However it is also possible that each contribution is individually small. In our plots this happens in fraction of points
of order $40\%$. Here even the gauge contribution to $\alpha$ might be dominant, especially if the Higgs is light.
We provide some analytical explanation of this fact in appendix \ref{sec:B}.

\subsubsection{Composite Right-handed Quarks}
\label{righthanded}

In the anarchic composite scenarios, since  symmetry breaking effects associated to the light generations are small,
the potential is dominated by the loops of third generation quarks. In \cite{MFVcomposite} it was however
shown that right handed light quarks could be strongly composite as long as they couple to singlets of
custodial symmetry as in the CHM$_5$ model (see also \cite{su2} for different realizations). In this case there are sizable
contributions to the potential also from the light generations.

\begin{figure}[t!]
\begin{center}
\subfigure
{\includegraphics[width=0.65\textwidth]{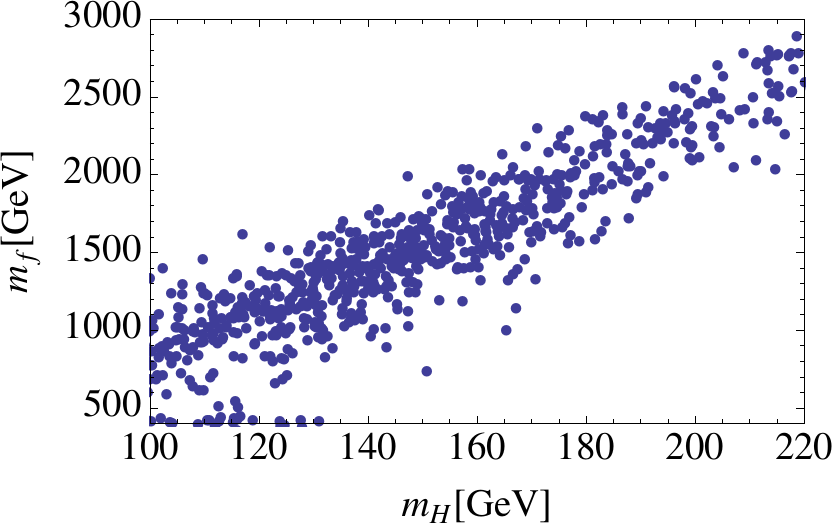}}\quad
\subfigure
{\includegraphics[width=0.65\textwidth]{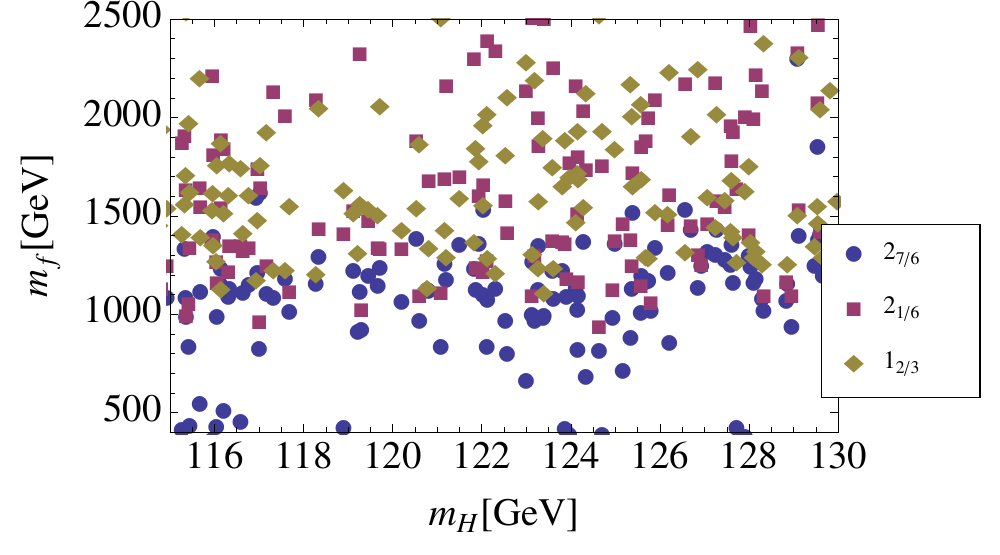}}
\caption{\label{fig:SO5MFVHiggs-all}\small Masses of fermionic partners as a function of the Higgs mass for $f=800$ GeV in CHM$_5$ with MFV.
Same parameters as in Fig. \ref{fig:SO5Higgs-all} are chosen.}
\end{center}
\end{figure}

This scenario is strongly motivated by flavor as it allows the realization of Minimal Flavor Violation in CHM.
This can be realized in CHM$_5$ if the strong sector has a flavor symmetry $U(3)_U \times U(3)_D$
respected by the right-handed mixings,
\begin{equation}\label{RHcomp}
\quad\Delta_{u_R}^{ij} \sim \delta^{ij}, \quad\Delta_{d_R}^{ij} \sim \delta^{ij}.
\end{equation}
so that the left-handed mixings are proportional to the SM Yukawas and are the only sources of
breaking of the flavor symmetries. Since the right-handed mixings are equal by the flavor symmetry to the ones of the third generation, the
contributions to the potential in particular from the up and charm quarks cannot be neglected.

The result of the scan is presented in Fig. \ref{fig:SO5MFVHiggs-all} with the same parameters as in Fig. \ref{fig:SO5Higgs-all}.
Despite the new contributions to the potential we find a correlation between the Higgs mass and the mass of the
lightest fermionic resonances similar to the one of the standard anarchic scenario. This can be easily understood.
The light quarks only contribute to the coefficient $\alpha$ because their contribution to  $\beta$
is small being proportional to the quark mass$^2$. After tuning the Higgs VEV, the Higgs mass only depends on $\beta$ (\ref{mhbeta})
that is dominated by the top Yukawa. To realize this configuration however a different correlation between the
left and right mixings is obtained.

\subsection{CHM$_{10}$}

\begin{figure}[t!]
\begin{center}
\subfigure
{\includegraphics[width=0.65\textwidth]{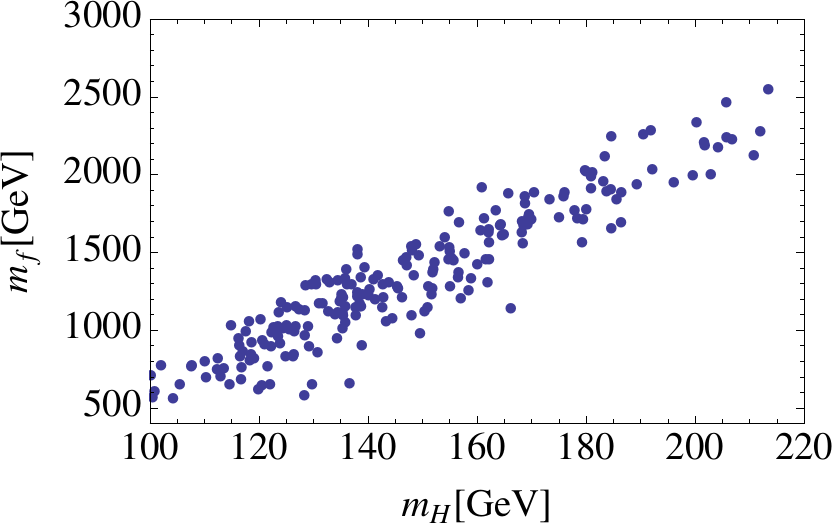}}\quad
\subfigure
{\includegraphics[width=0.65\textwidth]{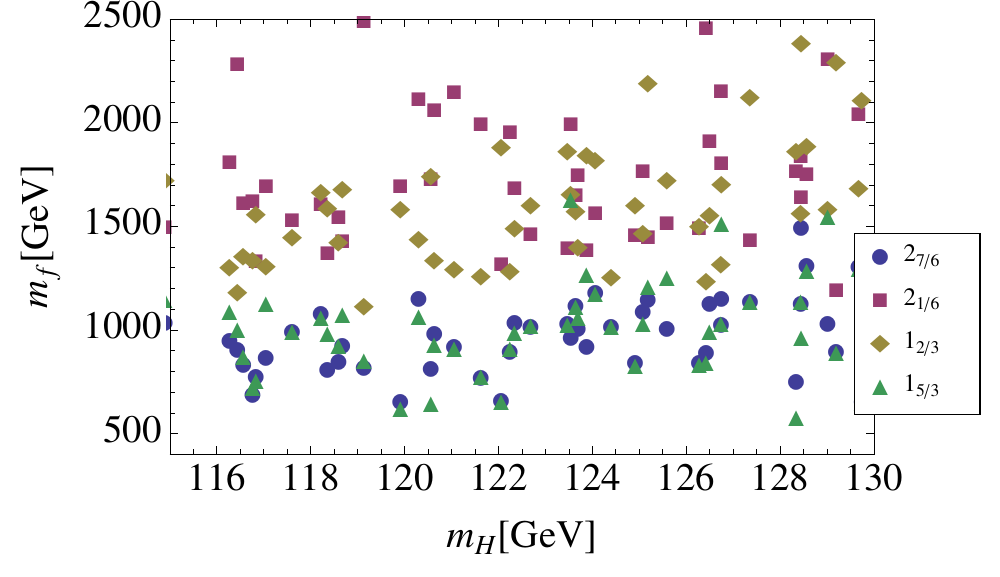}}
\caption{\label{fig:10Higgs-all}\small Higgs and fermionic partners masses in CHM$_{10}$ for $f=800$ GeV.
The mass parameters are chosen (in TeV): $1<\Delta_{t_L, t_R}<5$, $0.5<m_{T,\tilde{T}}<3$, $-2< m_{Y_T}< .5$ and $3<Y_T<6$.}
\end{center}
\end{figure}

Next we consider the model CHM$_{10}$ where the composite fermions are in the {\bf 10} of $SO(5)$.
This was originally studied in \cite{custodian} and in a different realization in \cite{serone}.  Under $SU(2)_L\times SU(2)_R$ the {\bf 10}
decomposes as a $({\bf 2},{\bf 2}) \oplus ({\bf 3},{\bf 1}) \oplus ({\bf 1},{\bf 3})$. Each chiral SM fermion couples
to a different ${\bf 10}_{2/3}$ of $SO(5)\times U(1)_X$.
The third generation quark lagrangian reads,
\begin{equation}
\label{CHM10}
\begin{split}
{\cal L}^{\rm CHM_{10}} &= \bar{q}^{el}_L i \slashed{D}^{el} q^{el}_L +  \bar{t}^{el}_R i \slashed{D}^{el} t^{el}_R\\
& + \Delta_{t_L} {\rm Tr}\left[\bar{q}_L \Omega \Psi_T \right] + \Delta_{t_R} {\rm Tr}\left[\bar{t}_R \Omega \Psi_{\widetilde T} \right] +h.c.\\
& + {\rm Tr}\left[\bar{\Psi}_T \left(i\slashed{D}^{\rho} - m_T \right)\Psi_T \right] + {\rm Tr}\left[\bar{\Psi}_{\widetilde T} \left(i\slashed{D}^{\rho} - m_{\widetilde T} \right)\Psi_{\widetilde T} \right] \\
&- Y_T \Phi^T \bar{\Psi}_{T,L} \Psi_{\widetilde T, R}\Phi - m_{Y_T}{\rm Tr}\left[ \bar{\Psi}_{T,L} \Psi_{\widetilde T, R} \right]+h.c.
\end{split}
\end{equation}
where $\Psi_{T,\tilde{T}}$ is a $5\times 5$ matrix corresponding to the adjoint rep of $SO(5)$. 
The lagrangian has structure analogous to eq.~\eqref{CHM5},  the presence of traces being needed to construct $SO(5)$ invariants.
Differently from CHM$_5$ the SM doublet needs to couple to a single composite fermion to generate the Yukawas of up and down sector.
As in CHM$_5$ the latter is not important for the potential and we will neglect it.

We performed a scan of the potential similarly to CHM$_5$ with $f=800~{\rm GeV}$. The main difference is
due to the fact that,
\begin{equation}
m_q^{\rm CHM_{10}} \sim  \frac 1 {\sqrt{2}} m_q^{\rm CHM_5}
\end{equation}
for the same choices of parameters so that larger elementary-composite mixings are typically required to reproduce the
top mass. Moreover because of different group theory factors (see \ref{aCHM10}) the size of the symmetry breaking
coupling $Y_T$ needs to be chosen larger than in CHM$_5$ to reproduce the electro-weak VEV and top mass.

The global scan scan  is shown in the upper plot in Fig. (\ref{fig:10Higgs-all}). In the lower plot we zoom in the low mass region.
The fermionic spectrum is reacher than CHM$_5$ containing a doublet ${\bf 2}_{1/6}$ a singlet ${\bf 1}_{2/3}$ a custodian ${\bf 2}_{7/6}$ a triplet ${\bf 3}_{2/3}$
and singlets ${\bf 1}_{5/3}$ (${\bf 1}_{-1/3}$). The last three are degenerate up to electro-weak symmetry breaking effects.
We note that in this model the first resonance tends to be lighter than in CHM$_5$ while the states
mixings with the SM fields are similar to that case. This is a consequence of the larger mixings required in this model that
forces the states that do not couple to SM fields to be lighter.

\section{Coset Space $SO(6)/SO(5)$}\label{sec:3}

The next simplest pattern of symmetry breaking relevant for CH is $SO(6)$ broken to $SO(5)$ originally introduced in \cite{singlet}.
There are 5 GBs  transforming in the vectorial representation of $SO(5)$,
that decomposes into a doublet and a singlet under the electro-weak symmetry group.
The coset space $SO(6)/SO(5)$ spanned by the GBs is the 5-sphere that can be conveniently parametrized by a unit vector of $SO(6)$,
\begin{equation}
\Phi=\sin \frac{\varphi}{f}\left( \frac {h_1}{\varphi},\,\frac {h_2}{\varphi},\,\frac {h_3}{\varphi},\,\frac {h_4}{\varphi},\,\frac {s}{\varphi},\,\cot \frac{\varphi}{f}\right)\,,~~~~~~~~
\varphi=\sqrt{\vec{h}^2+s^2}
\end{equation}
After electro-weak symmetry breaking one can choose the unitary gauge $h_1=h_2=h_3=0$ so that the physical degrees of freedom are $h=h_4$
and the singlet $s$.  The following parametrization will be convenient,
\begin{equation}
\begin{split}
h&=\varphi \cos\frac {\psi}f\\
s&=\varphi\sin\frac {\psi} f
\end{split}\quad \to \quad \Phi=\left(0,0,0,\sin\frac {\varphi}f \cos\frac {\psi}f,\sin\frac {\varphi}f\sin\frac {\psi}f,\cos\frac {\varphi}f\right)
\end{equation}
where $\varphi$ is a scalar and $\psi$ a pseudo-scalar under CP.  The kinetic terms correspond to a 2-sphere with standard metric,
\begin{equation}
ds^2= d\varphi^2 + \sin^2 \frac {\varphi}{f}\, d\psi^2
\label{2sphere}
\end{equation}

\subsection{Gauge Sector}

The gauge lagrangian has the same structure as  eq. (\ref{gaugeminimal}). The low energy
action for the Higgs and gauge fields is given by,
\begin{equation}\label{gauge6}
\frac{f^2}{2} (D_\mu \Phi)^T(D^\mu \Phi) = \frac 1 2 (\partial \varphi)^2 + \frac 1 2 \sin^2  \frac {\varphi} f
(\partial\psi)^2 + \frac{g^2f^2}{4}\sin^2\frac{\varphi}{f}\cos^2\frac{\psi}{f} W^2 + \cdots
\end{equation}
from which we identify the electro-weak VEV as,
\begin{equation}
v = f \sin \frac {\varphi}{f} \cos \frac {\psi}{f}
\label{vso6}
\end{equation}
As far as the gauge sector is concerned, on any vacuum one can always define a CP symmetry that is preserved.

\subsection{Fermionic Sector}

The simplest option for the composite fermions is that they are in a vector representation of $SO(6)$ and for simplicity we will focus on this here.
The fermionic action has the same structure as eq. (\ref{CHM5}). The embedding of the up quarks in the {\bf 6} reads
\begin{equation}\label{embed}
q_L \to \frac{1}{\sqrt{2}} \left(\begin{array}{c}
b_L\\ -i b_L \\ t_L \\ i t_L \\ 0 \\0
\end{array} \right),\quad t_R \to \left(\begin{array}{c}
0\\ 0 \\ 0 \\ 0 \\ e^{i\delta} \cos\theta t_R \\ \sin\theta t_R
\end{array} \right)
\end{equation}
and similarly for the down quarks. Note that $t_R$ can be coupled both to the fifth or sixth component of the vector, with complex phase $\delta$.
Since the action for the composite fields is (non-linearly) invariant under $SO(6)$, through an $SO(2)$ rotation of the fifth and sixth component
(under which the $SO(4)$ generators do not transform and the singlet shifts) and a phase redefinition of $t_R$ one can choose a basis where $\delta=\pi/2$.
In this basis it is manifest that the couplings of the top respects the CP symmetry \cite{cthdm},
\begin{equation}
\Psi \to \Omega_0\, (-i \gamma_2\gamma_0) \Psi^*\,,~~~~~~~~~~~~\Omega_0={\rm Diag}[1\,,-1\,,1\,,-1\,,-1\,,1]
\end{equation}
As in the CHM$_5$ CP could be explicitly broken by the phases of the strong sector but we will not consider this possibility here.

There are two special values of $\theta$. For $\theta=\pi/4$ the mixing of $t_R$ preserves the $SO(2)$ subgroup of $SO(6)$
which rotates the fifth and sixth component, simultaneously with a phase transformation of $t_R$.
In this limit the singlet is an exact GB and it will remain massless if the only contribution
to the potential is due to coupling to SM quarks\footnote{A tiny mass would be generated at 2-loops through anomalies with SM fields. In this case the singlet would be an electro-weak axion ruled out experimentally \cite{singlet}.}. $\theta=\pi/2$ is also special because the mixings leave an unbroken $Z_2$ symmetry under which $\psi$ changes sign. This guarantees the stability of the pseudo-scalar that could be used as dark matter candidate \cite{Frigerio:2012uc}.

\subsection{Higgs and Singlet Mass}

Let us now turn to the potential. Obviously the CP invariant vacuum $\psi=0$ is an extremum of the potential.
If $\psi$ acquires a VEV because $\psi=0$ is a maximum one might think that CP is spontaneously broken.
This is not the case however, at least in the limit where only the contribution of the top is included, see appendix \ref{sec:A}.
Inspecting the potential one can prove that if $\psi=0$ is a maximum, the minimum of the potential corresponds to,
\begin{equation}
\sin \frac {\varphi} f=1\,.
\end{equation}
In this vacuum the role of the fluctuations of $\psi$ and $\varphi$ are exchanged and  a different definition of CP exists that leaves the vacuum
invariant. One can check that this vacuum is physically equivalent to the one with $\psi=0$
so without loss of generality we will assume this configuration to be realized. We conclude that CP can be neither explicitly nor spontaneously broken
by the coupling to the top. A small breaking of CP could be induced from the contributions to the potential of the lighter generations neglected here.

\begin{figure}[t!]
\begin{center}
\subfigure
{\includegraphics[width=0.48\textwidth]{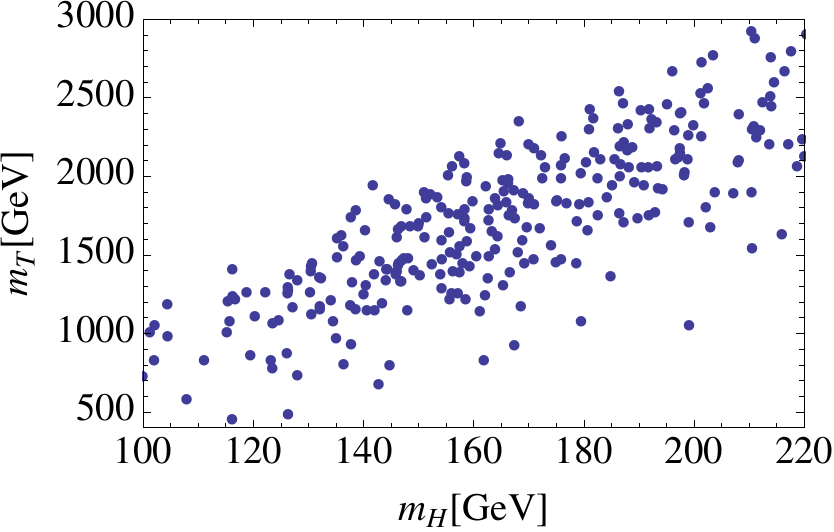}}\quad
{\includegraphics[width=0.48\textwidth]{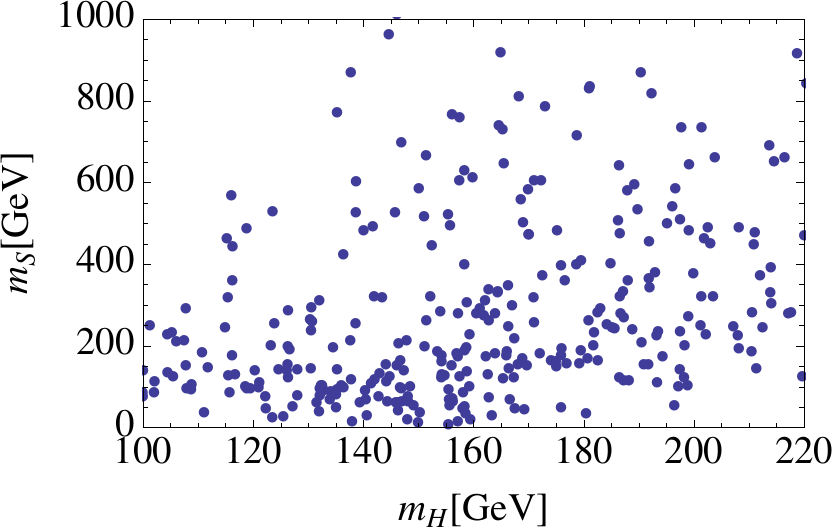}}
\subfigure
{\includegraphics[width=0.6\textwidth]{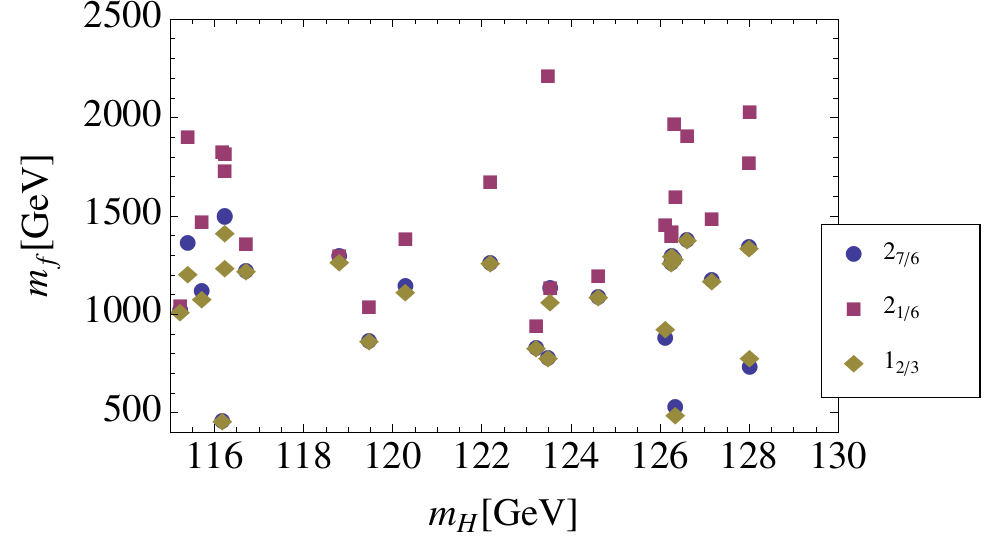}}
\caption{\label{fig:SO6Higgs-all} \small Masses of Higgs, singlet and fermionic partners in CHM$_6$ for $f=800$ GeV.
The six fermionic parameters are varied between $0.3$ and $4$ TeV with the mixing elementary-composite $\Delta_{t_L,t_R}/m_{T,\tilde{T}}<3$.
The gauge contribution corresponds to $f_1=f_2=\sqrt{2}f$  and $g_\rho=3$.
Up left Figure: Higgs vs. lightest fermionic state. Up right Figure: Higgs vs. singlet mass. Below masses of lightest singlet, doublet and custodian in the low
Higgs mass region.}
\end{center}
\end{figure}

The computation of the potential is similar to CHM$_5$, some details are in \ref{appA}. As far as Higgs mass is concerned
the formulas are identical to that case, the main difference being that in $SO(6)/SO(5)$, for equal choice of parameters one finds,
\begin{equation}
m_q^{\rm CHM_6} \sim m_q^{\rm CHM_5} \sin\theta.
\end{equation}
This means that to realize the top mass larger mixings of the elementary fields are necessary than in CHM$_5$, unless $\theta \sim \pi/2$.

We performed an unconstrained scan over the parameters of the model including the angle $\theta$. The range of parameters is chosen
as in CHM$_5$. In first of Figs. \ref{fig:SO6Higgs-all} we present a scan of points of parameter space with the correct EWSB and top mass for $f=800$.
We find roughly the same correlation between the Higgs mass and the mass of the lightest fermionic partners as in the $SO(5)/SO(4)$ model.
This is expected because the Higgs mass is controlled by $y_t^2$. The larger spread of masses can be understood as due to fact that larger mixings
are necessary if $\theta< \pi/2$.
Note that the lightest state in this case is often a singlet. This can be understood from the fact that contrary to CHM$_5$ one fermion singlet combination
does not mix with the SM fields so its mass is lower similarly to one of the ${\bf 2}_{7/6}$ doublet in CHM$_5$.

The mass of the CP odd singlet is typically heavier than the one of the Higgs. This is shown in second Fig. \ref{fig:SO6Higgs-all} .
This agrees with the fact that while the Higgs mass is controlled by the tuned
electro-weak VEV the mass of the singlet follows naturalness. However, if $\theta$ is close to $\pi/4$ the singlet becomes
an approximate GB and small values of the mass are then obtained. For $\theta=\pi/2$ the Higgs mass is identical to CHM$_5$ while
the singlet is heavy.

\subsection{Modified Couplings}

The low energy interactions of the Higgs and singlet to gauge fields and fermions can be parametrized similarly to Ref. \cite{doublehiggs} as,
\begin{equation}\label{low-energy}
\begin{split}
{\cal L}&=\frac 1 2 (\partial_\mu h)^2 +\frac 1 2 (\partial_\mu s)^2- V(h,s)+ \frac {v^2}{4} {\rm Tr} \left(D_\mu \Sigma^\dagger D^\mu \Sigma\right)\left[1+2 a_h \frac h v+ b_h \frac {h^2}{v^2}+b_s \frac {s^2}{v^2}\dots \right] \\
&-m_i \bar{\psi}_{Li}\Sigma\left(1+ c_h \frac h v+\dots\right)\psi_{Ri}-m_i \bar{\psi}_{Li}\left( c_s \frac s v+\dots\right)\psi_{Ri}+h.c.
\end{split}
\end{equation}
where $\Sigma=\exp(i\chi^a \sigma^a/v)$ contains the GBs eaten by $W$ and $Z$ bosons and we suppressed terms that violate CP
since the vacuum we are considering respects the symmetry.

\begin{figure}[t!]
\begin{center}
{\includegraphics[width=0.48\textwidth]{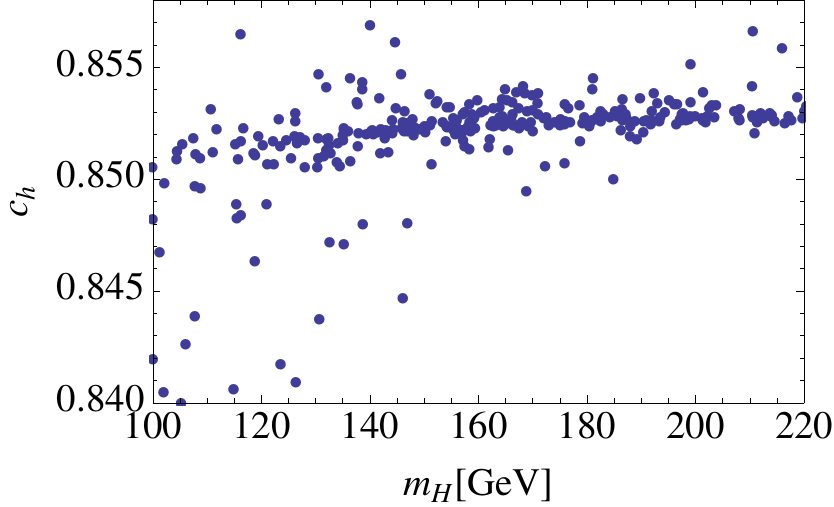}}\quad
\subfigure
{\includegraphics[width=0.48\textwidth]{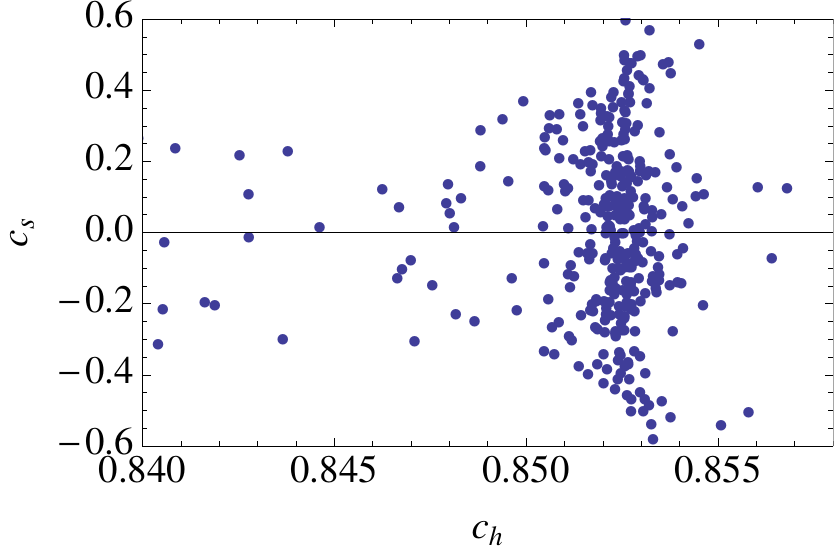}}\quad
\caption{\label{fig:SO6couplings} \small Coupling of the scalars to fermions including wave-function normalization effects. On the left
Higgs coupling to the top vs. Higgs mass. On the right the Higgs vs. singlet coupling.}
\end{center}
\end{figure}

The couplings of $a$ and $b$ are obtained expanding eq. (\ref{vso6}) around the vacuum
and taking into account the kinetic terms  (\ref{2sphere}),
\begin{equation}
a_h=\sqrt{1-\xi}\,,~~~~~ b_h=1- 2\xi\,,~~~~~~ b_s=1
\end{equation}
The coupling to fermions can be extracted from the effective action in \ref{sec:A} that contains,
\begin{equation}
\frac{ M_1^u(0)}{\sqrt{\Pi_{t_L}(0)\,\Pi_{t_R}(0)}} \bar{t}_L t_R s_\varphi c_\psi(i c_\theta s_\varphi  s_\psi+ s_\theta c_\varphi) + h.c.
\label{mtso6}
\end{equation}
where we abbreviate $\sin x/f\equiv  s_x$. Neglecting wave-function normalization effects
(arising from the Higgs dependence in $\Pi_{t_L}$ and $\Pi_{t_R}$) we find,
\begin{equation}
c_h= \frac {1- 2 \xi}{\sqrt{1-\xi}}\,~~~~~~~~
c_s=i  \sqrt{\frac {\xi} {1-\xi} }
\cot\theta
\end{equation}
In Fig. \ref{fig:SO6couplings} we compute the exact couplings to the fermions for the points of Fig. \ref{fig:SO6Higgs-all}.
On the left we see that on the point that realize the SM vacuum the effect of wave-function normalization is small even
for a light Higgs where some light partners exist. There is no correlation between the singlet coupling and the Higgs mass.

The coupling of the singlet to fermions could be in principle larger than the SM Higgs coupling,
however this requires a small value of $\theta$ where it is more difficult to obtain the top mass. Indeed, as shown in the second Fig. \ref{fig:SO6couplings},
in our sample it is always smaller. The couplings of the singlet vanish for $\theta=\pi/2$ as required by the $Z_2$ symmetry $\psi\to-\psi$.
Note that a possible phase in $M_1^u$, arising if the IR parameters $Y_T$ and $M_{Y_T}$ are complex
does not affect the couplings since it is reabsorbed into the top mass. This shows that the CP violating phase
of the strong sector does  not induce CP violation in the scalar couplings.

At loop level the Higgs and singlet will also couple SM gauge bosons. For gluons we parametrize,
\begin{equation}
{\cal L}_{ggh}=\frac{\alpha_s}{12\pi} g_{hGG} \frac{h}{v} G_{\mu\nu}^a G_{\mu\nu}^a,\quad\quad {\cal L}_{ggs}=\frac{\alpha_s}{16\pi} g_{sGG} \frac{s}{v} G_{\mu\nu}^a \tilde{G}_{\mu\nu}^a,
\end{equation}
where $\tilde{G}_{\mu\nu}^a=\epsilon_{\mu\nu\rho\sigma}{G}^{a\,\rho\sigma}$.
In the limit where Higgs and singlet are lighter than the top
but heavier than light quarks the coupling is determined by the top couplings above.
Following \cite{azatov-galloway} we derive,
\begin{equation}
\begin{split}
g_{hGG}&=v\left.\frac{\partial}{\partial \varphi} \log \det {\cal M}(\varphi,\psi) \right|_{\rm VEV}= \frac{1-2\xi}{\sqrt{1-\xi}}\\
g_{sGG}&=v\left.\frac{\partial}{\partial \psi} \log \det {\cal M}(\varphi,\psi) \right|_{\rm VEV}= \sqrt{\frac {\xi} {1-\xi} }\cot\theta,
\end{split}
\end{equation}
where ${\cal M}(\varphi,\psi)$ is the fermionic mass matrix. The coupling of the Higgs is the same as in CHM$_5$. The coupling of the singlet breaks the shift symmetry but does not vanish when $\theta=\pi/4$ where the Goldstone symmetry is recovered. This is due to fact that the $U(1)$ rotation of $t_R$ is anomalous.

\section{Discussion}
\label{sec:4}

In this note we have extended the study of the Higgs mass in composite Higgs models with partial compositeness
initiated in Ref. \cite{4dcomposite}. We have considered different fermionic setups for $SO(5)$ and studied
also the potential for $SO(6)/SO(5)$ where an extra CP odd scalar is present in the spectrum.
Along the way we have also clarified various CP properties of these theories. For example in
$SO(6)/SO(5)$ CP can neither be explicitly nor spontaneously broken by the couplings of the top
so we expect the coupling of the singlet to be CP preserving.

The main result is that a light Higgs, hinted by recent experimental results, typically requires
fermionic partners lighter than TeV. This represents a great opportunity for the LHC that will be able
to discover or exclude these fermionic partners in the early stages of the 14 TeV run, or even in the present run
under favorable circumstances. While not a general property of CHM in all the models considered, from eqs. (\ref{mhbeta}), (\ref{betanatural})
 the Higgs mass scales as,
\begin{equation}
m_h\sim \sqrt{\frac {N_c}2} \frac {y_t}{\pi} \frac {\Lambda}{f}\,v
\end{equation}
where $\Lambda$ is the cutoff of the quadratic divergence associated to the top Yukawa $y_t$. The cut-off is physically
represented by the lighter resonances, though not necessarily the lightest.
The necessity of light fermionic states can be understood in a model independent way from
naturalness of the theory. With a light Higgs, the unavoidable quadratic divergence of the Higgs mass generated by the top
Yukawa must be cut-off at a scale around TeV if a tuning of order 10\% is allowed (measured as $v^2/f^2$).
This demands new fermionic states to saturate the Higgs potential at a scale lower than spin-1 resonances.
In composite Higgs models other contributions to the potential exist but at least with the fermionic
representations considered here the Higgs mass is determined by the loops associated with the top Yukawa.
If contributions larger than that control the Higgs mass, naturalness would require even lighter states. Therefore our estimates
on the mass of the resonances should be considered as an upper bound.

We can  draw a parallel with supersymmetry:
\begin{figure}[ht!]
\begin{center}
{\includegraphics[width=.8\textwidth]{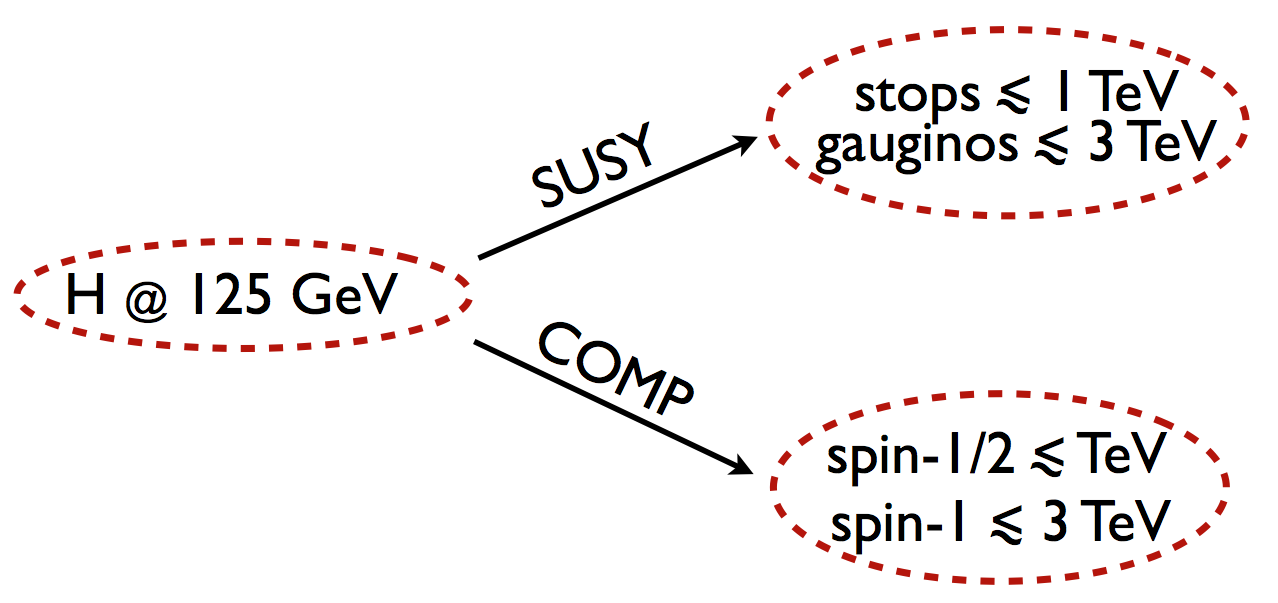}}\quad
\end{center}
\end{figure}

In  supersymmetry the naturalness of the electro-weak scale requires most minimally light stops
(below TeV) and possibly heavier winos and binos to cut-off the gauge loops, while other partners could be heavier and
even beyond the LHC reach.  This spectrum, dubbed natural SUSY, is to date the most compelling scenario for supersymmetry \cite{susy-natural}.
Bounds on light stops will soon reach $\sim 500$ GeV, a figure similar to the bound on fermionic partners in composite models.
If naturalness is a good guide, either new colored scalars or colored fermions should be soon discovered at the LHC.

\subsection*{Acknowledgments}
We are grateful to Stefania De Curtis for collaboration on related subjects.
We would like to thank Roberto Contino, Giuliano Panico, Riccardo Rattazzi and
Andrea Wulzer for stimulating discussions. We also thank GP and AW for sharing with us their results \cite{discrete2} before publication.

\appendix


\section{Basic Formulas}
\label{appA}

In this appendix we collect the basic formulas of the models studied in this paper. More details can be found in \cite{4dcomposite}.

For the gauge sector we have, up to electro-weak symmetry breaking effects,
\begin{equation}
\begin{split}
&m^2_{\rho}= \frac{g_\rho^2 f^2_1 }{2}\,,~~~~
m^2_{a_1}= \frac{g_\rho^2(f_1^2+f_2^2)}{2}\,,~~~~
f^2= \frac{f_1^2f_2^2}{f_1^2+f_2^2} \\
&\frac 1 {g^2}= \frac{1}{g_0^2}+\frac 1 {g_\rho^2}\,,~~~~~~~~~~~~~~
\frac{1}{g'^2}= \frac{1}{g_{0Y}^2}+\frac{1}{g_\rho^2}+\frac{1}{g_{\rho_X}^2},
\end{split}
\end{equation}
where $f$ is the decay constant of the GB Higgs and $g$, $g'$ are the SM couplings. $m_\rho$ and $m_{a_1}$ are the masses of resonances of $SO(4)$ and $SO(5)/SO(4)$, while $g_\rho$ and $g_{\rho_X}$ are the couplings of the composite resonances of $SO(5)$ and $U(1)_X$. The mass of $\rho_X$ is a free parameter.

\subsection{$SO(5)/SO(4)$}

The effective action for SM fermions in CHM$_5$ and CHM$_{10}$ obtained integrating out the strong sector takes the form  \cite{custodian},
\begin{equation}
\begin{aligned}
{\cal L}_{\rm eff} &=
\bar q_L \slashed{p} \left[ \Pi_0^q(p^2)
 + \frac{s^2_h}{2} \left( \Pi_1^{q1}(p^2)\, \widehat H^c \widehat H^{c\dagger}
 +  \Pi_1^{q2}(p^2)\, \widehat H \widehat H^\dagger \right) \right] q_L \\
& +\bar u_R \slashed{p} \left( \Pi_0^u(p^2) + \frac{s^2_h}{2}\, \Pi_1^u (p^2)\right) u_R
 +\bar d_R \slashed{p} \left( \Pi_0^d (p^2)+ \frac{s^2_h}{2}\, \Pi_1^d (p^2)\right) d_R  \\
&+ \frac{s_hc_h}{\sqrt{2}} M_1^u (p^2)\,\bar q_L \widehat H^c u_R
 + \frac{s_hc_h}{\sqrt{2}} M_1^d (p^2)\,\bar q_L \widehat H d_R + h.c. \,  .
\end{aligned}
\label{fermgen}
\end{equation}
The 1-loop effective potential due to the top loop is given by,
\begin{equation}
\label{pot_fer}
V(h)_{fermions}= -2N_c \int\frac{ d^4p}{(2\pi)^4}\left[\ln \Pi_{b_L} + \ln \left(p^2 \Pi_{t_L}\Pi_{t_R} - \Pi^2_{t_Lt_R} \right)\right]
\end{equation}
where,
\begin{eqnarray}
\Pi_{t_L}&=& \Pi^{q}_0 + \frac{\sin^2 (h/f)}{2} \Pi_1^{q_1}\,,~~~~~~~~~~~~~~~
\Pi_{b_L}= \Pi^{q}_0 + \frac{\sin^2 (h/f)}{2} \Pi_1^{q_2}\nonumber\\
\Pi_{t_R}&=& \Pi_{0}^{u} +\frac{\sin^2 (h/f)}{2} \Pi_1^{u}\,,~~~~~~~~~~~~~~\Pi_{t_Lt_R} = \frac{\sin (h/f)\cos (h/f)}{\sqrt{2}}M^{u}_1.
\end{eqnarray}
For $s_h\ll1$ one can approximate,
\begin{equation}
V(h)_{fermions}\approx  \alpha s_h^2 -\beta s_h^2 c_h^2
\label{apppot}
\end{equation}
where
\begin{eqnarray}
\label{pot_param}
\alpha &=& -N_c \int \frac{d^4p}{(2\pi)^4} \left[ \frac{\Pi_1^{q_1}}{\Pi_0^{q}} + \frac{\Pi_1^{u}}{\Pi_0^{u}}\right]\nonumber \\
\beta &=& -N_c \int \frac {d^4p}{(2\pi)^4} \frac{(M_1^{u})^2}{p^2 (\Pi_0^{q}+s_h^2/2 \Pi_1^{q_1}) (\Pi_0^{u}+s_h^2/2 \Pi_1^{u})}
\end{eqnarray}
The gauge loops also contribute to $\alpha$.

In our model the form factors can be expressed in terms of the functions,
\begin{equation}
\label{bido-sing}
\begin{aligned}
\widehat{\Pi}[m_1,m_2, m_3]&= \frac{\left(m_2^2+m_3^2-p^2\right)\,\Delta^2}{p^4 - p^2 (m_1^2+m_2^2+m_3^2) +m_1^2m_2^2},\\
\widehat{M}[m_1, m_2, m_3]&=- \frac{m_1 m_2 m_3\,\Delta^2}{p^4 - p^2 (m_1^2+m_2^2+m_3^2) +m_1^2m_2^2}.
\end{aligned}
\end{equation}
as follows:

\subsubsection{CHM$_5$}
\label{appCHM5}

Including only the top the form factors are given by,
\begin{equation} \label{CHM5-Piq}
\begin{split}
\Pi_0^q &= \frac{1}{y_{t_L}^2}+ \widehat\Pi_0^{q_L} \, , \\
\Pi_0^u &=\frac{1}{y_{t_R}^2}+ \widehat\Pi_0^{u_R} + \widehat\Pi_1^{u_R} \, , \\
\end{split} \qquad
\begin{split}
&\Pi_1^{q_1}= \widehat\Pi_1^{q_L}\, , \\
&\Pi_1^{q_2}= 0\, ,\\
&\Pi_1^{u}= -2\, \widehat\Pi_1^{u_R}\
\end{split} \qquad
\begin{split}
M_1^u &= \widehat M_1^{u}\, , \\[0.05cm]
\end{split}
\end{equation}
where
\begin{eqnarray}
\label{SE_5y}
\widehat{\Pi}^{q_L}_0&=&\widehat{\Pi}[m_T, m_{\widetilde{T}}, m_{Y_T}]\,,~~~~~~~\widehat{\Pi}^{q_L}_1=\widehat{\Pi}[m_T, m_{\widetilde{T}}, m_{Y_T}+Y_T]-\widehat{\Pi}[m_T, m_{\widetilde{T}}, m_{Y_T}], \nonumber \\
\widehat{\Pi}^{u_R}_0&=&\widehat{\Pi}[m_{\widetilde{T}}, m_T , m_{Y_T}]\,,~~~~~~~\widehat{\Pi}^{u_R}_1=\widehat{\Pi}[m_{\widetilde{T}}, m_T , m_{Y_T}+Y_T]-\widehat{\Pi}[ m_{\widetilde{T}}, m_T , m_{Y_T}], \nonumber \\
\widehat{M}^{u}_0&=& \widehat{M}[m_T, m_{\widetilde{T}}, m_{Y_T}]\,,~~~~~~\widehat{M}^{u}_1=\widehat{\Pi}[m_T, m_{\widetilde{T}}, m_{Y_T}+Y_T]-\widehat{\Pi}[m_T, m_{\widetilde{T}}, m_{Y_T}].\nonumber\\
\end{eqnarray}
and to match with the lagrangian (\ref{CHM5}) $\Delta_{t_{L},t_R}=y_{t_{L},t_{R}} \Delta$.

Up to electro-weak symmetry breaking effetts the masses of the $\bf{2}_{1/6}$, $\bf{2}_{7,6}$ and $\bf{1}_{2/3}$ fermions are given by the zeros of $\Pi_0^q$, the poles of $\Pi_0^q$ and the zeros of $\Pi_0^u$ respectively.

\subsubsection{CHM$_{10}$}
\label{aCHM10}

In this case one finds,
\begin{equation}
\label{CHM10-Piq}
\begin{split}
\Pi_0^q &= \frac{1}{y_{t_L}^2}+ \widehat\Pi^{q_L}_0+\frac{1}{2}\, \widehat\Pi^{q_L}_1 \, , \\
\Pi_0^u &= \frac{1}{y_{t_R}^2}+ \widehat\Pi^{u_R}_0 \, , \\
\end{split} \qquad
\begin{split}
\Pi_1^{q_1}&= -\frac{1}{2}\, \widehat\Pi^{q_L}_1 \, , \\
\Pi_1^{q_2}&= -\widehat\Pi^{q_L}_1 \, , \\
\Pi_1^{u}  &= \frac{1}{2}\, \widehat\Pi^{u_R}_1 \, ,
\end{split} \qquad
\begin{split}
M_1^u &= \frac{1}{2\sqrt{2}}\, \widehat M^{u}_1  \, , \\[0.05cm]
\end{split}
\end{equation}
where
\begin{eqnarray}
\label{CHM10PI}
\widehat{\Pi}^{q_L}_0&=&\widehat{\Pi}[m_T, m_{\widetilde{T}}, m_{Y_T}]\,,~~~~~~~\widehat{\Pi}^{q_L}_1=2 \widehat{\Pi}[m_T, m_{\widetilde{T}}, m_{Y_T}+Y_T/2]-2\widehat{\Pi}[m_T, m_{\widetilde{T}}, m_{Y_T}] ,\nonumber \\
\widehat{\Pi}^{u_R}_0&=&\widehat{\Pi}[m_{\widetilde{T}}, m_T , m_{Y_T}]\,,~~~~~~~\widehat{\Pi}^{u_R}_1=2 \widehat{\Pi}[m_{\widetilde{T}}, m_T , m_{Y_T}+Y_T/2]- 2\widehat{\Pi}[ m_{\widetilde{T}}, m_T , m_{Y_T}], \nonumber \\
\widehat{M}^{u}_0&=& \widehat{M}[m_T, m_{\widetilde{T}}, m_{Y_T}]\,,~~~~~~\widehat{M}^{u}_1=2 \widehat{\Pi}[m_T, m_{\widetilde{T}}, m_{Y_T}+Y_T/2]-2 \widehat{\Pi}[m_T, m_{\widetilde{T}}, m_{Y_T}].\nonumber\\
\end{eqnarray}
The masses of the $\bf{2}_{1/6}$, $\bf{2}_{7,6}$ $\bf{1}_{2/3}$ and $(\bf{1}_{5/3},\bf{1}_{-1/3},\bf{3}_{2/3})$ fermions are given by the zeros of $\Pi_0^q$, the poles of $\Pi_0^q$, the zeros of $\Pi_0^u$ and the poles $\Pi_0^u$ respectively.

\subsection{$SO(6)/SO(5)$}
\label{sec:A}

With composite fermions in the $\bf{6}$ the effective lagrangian for the SM fermions reads,
\begin{eqnarray}
\label{fer_eff2}
{\cal L}_{{\rm eff}} &=&\ \!
 \bar{q}_L  \slashed{p} \left( \widehat{\Pi}^{q_L}_0(p)
  + \frac{1}{2}\frac{s_\varphi^2}{\varphi^2} \widehat{\Pi}_1^{q_L}(p) H^c H^c \right)\!q_L\nonumber\\
  &+& \! \bar{u}_R \slashed{p} \left( \widehat{\Pi}_0^{u_R}(p) + s_\theta^2 \widehat\Pi_1^{u_R}(p) + \left[ s_\varphi^2 ( c_\theta^2 s_\psi^2 -s_\theta^2) \right]\widehat\Pi_1^{u_R}(p)  \right)u_R \nonumber \\
 &+& \frac{\widehat{M}_1^{u}(p)}{\sqrt{2}} \frac{s_\varphi}{\varphi}\bar{q}_L H^c \left( i c_\theta s_\varphi s_\psi + s_\theta c_\varphi \right)u_R
 + h.c.
\end{eqnarray}
where the form factors are,
\begin{equation} \label{CHM5-Piq}
\begin{split}
\Pi_0^q &= \frac{1}{y_{t_L}^2}+ \widehat\Pi_0^{q_L} \, , \\
\Pi_0^u &=\frac{1}{y_{t_R}^2}+ \widehat\Pi_0^{u_R} +s_\theta^2\, \widehat\Pi_1^{u_R} \, , \\
\end{split} \qquad
\begin{split}
&\Pi_1^{q_1}= \widehat\Pi_1^{q_L}\, , \\
&\Pi_1^{q_2}= 0\, ,\\
&\Pi_1^{u}= -2\, \widehat\Pi_1^{u_R}\
\end{split} \qquad
\begin{split}
M_1^u &= \widehat M_1^{u}\, , \\[0.05cm]
\end{split}
\end{equation}
in terms of the building blocks (\ref{SE_5y}).

The potential has now the form,
\begin{equation}
\begin{split}
V(\varphi,\psi) &\simeq c_1 \sin^2 \varphi \cos^2\psi  + c_2 \sin^2\varphi\big(\sin^2\theta-\cos^2\theta \sin^2\psi \big)\\
 &-c_3 \sin^2\varphi\cos^2\psi \big[\cos^2\theta \sin^2\varphi \sin^2\psi + \sin^2\theta\cos^2\varphi\big]
\end{split}
\end{equation}
where,
\begin{equation}
\begin{split}
&c_1 = -N_c \int\frac{d^4p}{(2\pi)^4} \frac{\Pi_1^{q_1}}{\Pi_0^{q}}, \quad c_2=-N_c \int\frac{ d^4p}{(2\pi)^4} \frac{\Pi_1^{u}}{\Pi_0^{u}},\\
&c_3=- N_c \int\frac{d^4p}{(2\pi)^4}\frac{(M_1^{u})^2}{(\Pi_0^{q}+s_h^2\, \Pi_1^{q_1}/2) (\Pi_0^{u}+s_h^2 s_\theta^2 \Pi_1^{u}/2)}
\end{split}\end{equation}

The gauge loops also contribute to  the coefficient $c_1$.
Two extrema of the potential are $\psi=0$ and $s_\varphi=1$. One can prove that other extrema of the potential
are saddle points so that one is the minimum while the other is the maximum.
Note that since the lagrangian is invariant under the transformations,
\begin{eqnarray}
s_\varphi s_\psi&\to& c_\varphi' \nonumber \\
c_\varphi &\to& s_\varphi' s_\psi' \nonumber \\
s_\theta &\to& c_\theta
\end{eqnarray}
one can choose without loss of generality the minimum at $\psi=0$.  The Higgs and singlet mass are then given by,
\begin{eqnarray}
m_h^2&=& \frac {-4 c_1 c_2-2 c_1^2/s_\theta^2+2(c_3^2-c_2^2) s_\theta^2}{c_3\,f^2}\nonumber \\
m_s^2 & =& \frac {c_1-(c_2+ c_3) s_\theta^2}{f^2\,s_\theta^2}c_{2\theta}
\end{eqnarray}

\section{Potential}
\label{sec:B}

Here we provide some analytical formulas for the potential.
The leading order contribution to the potential from gauge loops (neglecting hyper-charge effects) is given by \cite{4dcomposite},
\begin{equation}
\label{explicit-pot-gauge}
V(h)_{gauge}\approx \frac{9}{4}  \frac{1}{16\pi^2} \frac{g_0^2}{g_\rho^2} \frac{m_{\rho}^4 \left(m_{a_1}^2-m_{\rho}^2\right)}{m_{a_1}^2-m_{\rho}^2(1+g_0^2/g_\rho^2)}\ln \left[\frac{m_{a_1}^2}{m_{\rho}^2(1+g_0^2/g_\rho^2)} \right]\sin^2 \frac h f.
\end{equation}
This formula agrees with our intuition in various ways. In the SM the quadratic divergence of the Higgs mass due to gauge loops reads,
\begin{equation}
\delta m_h^2 =\frac {9\, g^2} {32\pi^2}  \Lambda^2
\end{equation}
where $\Lambda$ is a momentum cut-off. Comparing with the formula above we see that the $SO(4)$ resonances
act as the cut-off of the gauge loops. More precisely we see that the quadratic divergence is cut-off at the mass of
triplet mixing with SM gauge fields whose mass is $m_\rho \sqrt{1+g_0^2/g_\rho^2}$.
The residual logarithmic sensitivity to the cutoff is regulated by the coset  resonances in $SO(5)/SO(4)$.

The expressions for the fermions (\ref{pot_param}) are more involved but the physics similar. To be definite let us consider CHM$_5$.
To gain some insight into these formulas we can expand around $\Delta_{t_L,t_R}=0$. To leading order we have,
\begin{eqnarray}
\alpha &=& -\frac {N_c} {16\pi^2}\int_0^\infty dt \left[\frac {(m_1^2-m_3^2)(m_3^2-m_2^2)[ \Delta_{t_L}^2m_Q^2 t (t+m_T^2)  - 2 \Delta_{t_R}^2 m_T^2 t (t+m_Q^2)]}{m_3^2(t+m_1^2)(t+m_2^2)(t+m_3^2)(t+m_4^2)}\right]+\dots \nonumber \\
\beta &=& \frac {N_c \Delta_{t_L}^2 \Delta_{t_R}^2}{64 \pi^2}\int_0^\infty dt \left[\frac{m_1 m_2 (m_1^2-m_3^2)(m_3^2-m_2^2)+m_3^2 Y_T^2}{Y_T\,m_3^2 (t+m_1^2)(t+m_2^2)}+
 \left(\begin{array}{c}
 m_1 \leftrightarrow m_3\\ m_2 \leftrightarrow m_4
  \end{array}\right)\right]^2+\dots
\nonumber \\
\end{eqnarray}
where $t\equiv - p^2$ and
we have expressed the integrals in terms of the physical masses of the singlets ($m_{1,2}$) and quadruplets
($m_{3,4}$). In the limit we are working each $SO(4)$ rep is degenerate. Note that the following relations hold in our model,
\begin{equation}
m_1\, m_2 = m_3\, m_4\,,~~~~~~~~~~~~~~y_t\approx \frac {\Delta_{t_L}\, \Delta_{t_R}}{m_1\, m_2}\frac {Y_T} f
\label{appmasses}
\end{equation}
For $\Lambda \ll m_i$ the integrals behave as,
\begin{eqnarray}
\alpha&=& \frac {N_c }{16\pi^2} \frac {(2 \Delta_{t_R}^2-\Delta_{t_L}^2)(m_1^2-m_3^2)(m_2^2-m_3^2)}{m_1^2 m_2^2 m_3^2}\Lambda^4\nonumber \\
\beta &=&  \frac {N_c\, y_t^2}{16\pi^2} f^2 \Lambda^2
\end{eqnarray}
From the low energy point of view $\alpha$ is quartically divergent. This contribution originates from the
Higgs dependence of the kinetic terms that behaves as the vacuum energy.
$\beta$ reproduces the quadratic divergence of the top in the SM with the appropriate coefficient.
The divergences are reduced to logarithmic above $m_1$ and $m_3$.
Both the singlet and the quadruplet are necessary to  regulate the integrals.
This is very similar to supersymmetry where the analogous cancellation is due to left and right stops.
The mixing terms in $\beta$ are also reminiscent of the $A-$terms.

The integrals above can be done analytically but the answer is not particularly illuminating. To get a more useful expression we can
decouple the heavy singlet and quadruplet by taking $m_{2,4}\to \infty$ according to eq. \eqref{appmasses}. To get a finite Yukawa we then have to scale $Y_T=\kappa \, \sqrt{m_2 m_4}$.  In this limit we can interpret $\kappa\,\sqrt{m_1 m_2}/f$ as the fermionic coupling of the strong sector.
We note that the integral of $\beta$ is finite in this limit while $\alpha$ in general requires also the second layer of resonances to
regulate a residual logarithmic divergence. We find,
\begin{equation}
\begin{split}
\beta =\frac {N_c y_t^2}{64 \pi^2 \kappa^4}&\times
\Big[\frac {(m_1^2-m_3^2)^3(m_1^2+m_3^2)^2-2 m_1 m_3(m_1^2-m_3^2)^3 \kappa^2+m_1^2 m_3^2 (m_1^4-m_3^4)\kappa^4}{m_1^2 m_3^2(m_1^2-m_3^2)}\, \\
&+\frac {4 m_1^2 m_3^2(m_1^4+m_3^4-m_1^2 m_3^2(2+\kappa^4) \log \frac {m_3}{m_1}}{m_1^2 m_3^2(m_1^2-m_3^2)}\Big] f^2
\end{split}
\end{equation}
For $m_1=m_3$ we find simply,
\begin{equation}
\beta= \frac {N_c\, y_t^2}{16\pi^2} f^2 m_1^2 \,.
\end{equation}
$\alpha$ is also calculable for this choice and both contributions (proportional to $\Delta_{t_L,t_R}^2$) vanish.
Deviating from this point one should be able to find solutions. Indeed in a sizable fraction of our samples
the coefficient $\alpha$ is small compared to the naive estimate because it is the sum of small terms
rather the cancellation of large terms. Other regions of parameters space however can also be found.

\end{document}